\documentclass[aps,prd,twocolumn,floatfix,nofootinbib,superscriptaddress]{revtex4-1}
\usepackage[colorlinks,linkcolor=blue,anchorcolor=blue,citecolor=blue,urlcolor=blue,breaklinks=true]{hyperref}
\usepackage[libertine]{newtxmath}
\usepackage{graphicx}
\usepackage{slashed}
\usepackage{amsfonts}
\usepackage{amsmath}
\usepackage{blindtext}
\usepackage{microtype}
\usepackage{textcomp}
\usepackage{comment}

\newcommand{\vect}[1]{\boldsymbol{#1}}

\newcommand{\sh}[1]{\slashed{#1}}
\newcommand{\da}{\downarrow}
\newcommand{\ua}{\uparrow}

\newcommand{\CDOT}{\mbox{$ \cdot $}}
\newcommand{\kslash}{\mbox{$\not \! k$}}

\begin{document}

\title{Pion and $\rho$ meson's unpolarized quark distribution functions from $q\bar{q}$ and all Fock-states within Dyson--Schwinger equations.}

\author{Chao Shi}
\email[]{cshi@nuaa.edu.cn}
\affiliation{Department of Nuclear Science and Technology, Nanjing University of Aeronautics and Astronautics, Nanjing 210016, China}

\author{Liming Lu}
\affiliation{Department of Nuclear Science and Technology, Nanjing University of Aeronautics and Astronautics, Nanjing 210016, China}

\author{Wenbao Jia}
\affiliation{Department of Nuclear Science and Technology, Nanjing University of Aeronautics and Astronautics, Nanjing 210016, China}

\begin{abstract}
We compute the twist-2 unpolarized quark parton distribution functions (PDFs) of the pion and the $\rho$ meson within the Dyson--Schwinger equations (DSEs) framework using the rainbow--ladder (RL) truncation. A new DSE for the spin-1 hadron's quark--quark correlation matrix is derived, from which the PDFs can be extracted. For the $\rho$ meson, we obtain for the first time within RL-DSEs the unpolarized PDFs corresponding to different helicity states. A pronounced difference is observed between the $|\Lambda|=1$ and $\Lambda=0$ cases, where $\Lambda$ denotes the meson helicity, leading to a nonvanishing and numerically sizable tensor-polarized PDF $f_{1LL}(x)$. We further compare these results with those obtained under a leading Fock-state ($|q\bar{q}\rangle$) truncation and find substantial deviations. This comparison demonstrates that the present RL-DSEs framework incorporates higher Fock-state contributions associated with gluonic degrees of freedom, which has a significant impact on the quark distributions.
\end{abstract}

\maketitle

\section{Introduction}

Quantum chromodynamics (QCD) is distinguished from Abelian gauge theories by the non-Abelian nature of its gauge group, which gives rise to gluon self-interactions and a rich infrared structure of the strong interaction. These features underpin two defining properties of QCD: asymptotic freedom at short distances and confinement at hadronic scales~\cite{Gross:1973id,Politzer:1973fx}. As a consequence, hadrons are strongly correlated systems in which gluonic degrees of freedom play an essential dynamical role and strongly affect the partonic structure of hadrons at low, nonperturbative momentum scales. In the era of high-precision parton imaging and with the EIC targeting gluon distributions \cite{LHeC:2020van,AbdulKhalek:2021gbh,Anderle:2021wcy}, it has become imperative to incorporate the impact of gluonic Fock components into calculations of hadronic quark parton distribution functions (PDFs) within nonperturbative QCD frameworks.

In recent years, several Euclidean-space DSE studies have begun to disentangle quark and gluon contributions to hadronic PDFs. In Ref.~\cite{Bednar:2018mtf}, a pioneering rainbow--ladder DSE calculation showed that quarks carry approximately 70\% of the pion's and kaon's light-front longitudinal momentum at a hadronic scale, implying that the remaining 30\% is carried by gluons. This result was later confirmed by direct calculations of the gluon PDFs of the pion and nucleon \cite{Freese:2021zne}, as well as by studies of the gluon contribution to the nucleon spin decomposition \cite{Tandy:2023zio}. Meanwhile, an independent line of DSE investigations explored the leading Fock-state ($|q\bar{q}\rangle$) light-front wave functions (LFWFs) of mesons \cite{Shi:2018zqd,Shi:2021taf,Shi:2021nvg,Shi:2023jyk}. It was observed that the resulting $q\bar{q}$-LFWFs satisfy $\langle q\bar{q}|q\bar{q}\rangle < 1$, indicating the presence of higher Fock components associated with gluonic degrees of freedom. The light $\rho$ and $\phi$ meson $q\bar{q}$ LFWFs were further tested through diffractive vector-meson production within the color-dipole model \cite{Shi:2021taf,Shi:2025mne}. Good agreement with HERA data at large $Q^2$ was achieved, confirming the dominance of the $q\bar{q}$ truncation in exclusive processes at high momentum transfer. We emphasize that the DSE studies mentioned above incorporate quark interactions mediated by modeled gluon propagators; it is therefore natural that they encode gluonic Fock components on the light front.

Recently, motivated by the physical insight of Ref.~\cite{Bednar:2018mtf}, we developed a new framework and applied it to the study of the pion's twist-3 distribution function $e(x)$. In Ref.~\cite{Bednar:2018mtf}, gluon dressing effects were included in the probing light-cone vertex, whereas in Ref.~\cite{Shi:2026wca} the corresponding resummation was absorbed into the pion's quark--quark correlation matrix. In the present work, we generalize the framework of Ref.~\cite{Shi:2026wca} to the vector-meson (spin-1 hadron) case. Our aim is to obtain the unpolarized quark PDF of the $\rho$ meson in a manner that effectively incorporates contributions from all Fock sectors. We also compare these full PDFs (including the pion case) with those obtained from $q\bar{q}$-LFWFs. A substantial deviation between the full and $q\bar{q}$-truncated PDFs is observed---potentially the largest reported in the literature. We note that the partonic structure of spin-1 targets attracts considerable interest \cite{Jaffe:1988up,Hoodbhoy:1988am,Miller:2013hla,Kumano:2020ijt,Kumano:2021fem,Kumano:2021xau,Kumano:2024fpr,Bacchetta:2000jk,Cotogno:2017puy}—particularly the deuteron, which is experimentally accessible. Although the $\rho$ meson is experimentally challenging, it provides a valuable theoretical laboratory as a highly relativistic bound state that can exhibit rich and novel partonic structure \cite{Ninomiya:2017ggn,Puhan:2023hio,Kaur:2024iwn,Zhang:2024plq,Liu:2025fuf,Tanisha:2025qda}.

This paper is organized as follows. In Sec.~\ref{sec:LFWF}, we briefly review our previous work on extracting $q\bar{q}$-LFWFs within the DSEs framework. In Sec.~\ref{sec:DSE}, we derive the DSE for the spin-1 hadron quark--quark correlation matrix, describe its numerical solution, and explain the extraction of the unpolarized quark PDFs. In Sec.~\ref{sec:results}, we report the resulting full PDFs and compare them with the contributions from the $|q\bar{q}\rangle$ sector. Finally, we summarize our findings in Sec.~\ref{sec:con}.

\section{$q\bar{q}-$contribution to unpolarized PDFs.}\label{sec:LFWF}

It has long been recognized that the $q\bar q$-LFWFs of mesons can be obtained from their BS wave functions~\cite{tHooft:1974pnl,Lepage:1980fj,Frederico:2011ws}. Based on this idea, in recent years we have extracted the $q\bar q$-LFWFs of pseudoscalar and vector mesons from their Euclidean-DSE based BS wave functions~\cite{Shi:2018zqd,Shi:2021taf,Shi:2021nvg,Shi:2023jyk}. These LFWFs can then be used to compute meson PDFs. Since the resulting PDFs include only contributions from the leading Fock sector, we refer to them as $q\bar q$-PDFs in following. In this section we briefly summarize the formalism and the main results.

In light front QCD, the leading Fock-state of a meson reads 
\begin{align}\label{eq:LFWF1}
	|M\rangle^{\{\Lambda\}} &= \sum_{\lambda,\lambda'}\int \frac{d^2 \vect{k}_T}{(2\pi)^3}\,\frac{dx}{2\sqrt{x\bar{x}}}\, \frac{\delta_{ij}}{\sqrt{3}} \nonumber  \\
	&\hspace{10mm} \Phi^{\{\Lambda\}}_{\lambda,\lambda'}(x,\vect{k}_T)\, b^\dagger_{f,\lambda,i}(x,\vect{k}_T)\, d_{g,\lambda',j}^\dagger(\bar{x},\bar{\vect{k}}_T)|0\rangle.
\end{align}
Here, the bracket $\{\Lambda\}$ is included only for vector mesons, with $\Lambda$ denoting the meson helicity. For pseudoscalar mesons, the entire bracketed $\{\Lambda\}$ should be dropped.
The $\Phi^{\{\Lambda\}}_{\lambda,\lambda'}$ is the $q\bar{q}$-LFWF of the considered meson, with quark (antiquark) carrying spin $\lambda$ ($ \lambda'$). The $\Lambda=0, \pm 1$ for vector meson, and $\lambda=\ua$ or $\da$, which will be denoted as $\ua=+$ and $\da=-$ for abbreviation in the following. The $i$ and $j$ are color indices. The $\vect{k}_T=(k^x,k^y)$ is the transverse momentum of the quark with flavor $f$, and $\bar{\vect{k}}_T=-\vect{k}_T$ for antiquark with flavor $g$. The longitudinal momentum fraction carried by quark is $x=k^+/P^+$, with $\bar{x}=1-x$ for antiquark. Light-cone four-vector of this paper takes the convention $A^{\pm} = \tfrac{1}{\sqrt{2}}(A^0 \pm A^3)$. 

The meson's $q\bar{q}$-LFWFs can be extracted from their covariant BS wave functions with \cite{Shi:2021taf} \footnote{This equation is derived in Minkowski space. In practice, we transform it to $x$-Mellin moment space, and perform a transformation to Euclidean space in order to match the Bethe--Salpeter wave functions obtained in Euclidean formulation.}
\begin{align}\label{eq:chi2phi}
	\Phi^{\{\Lambda\}}_{\lambda,\lambda'}(x,\vect{k}_T)&=-\frac{1}{2\sqrt{3}}\int \frac{dk^- dk^+}{2 \pi} \delta(x P^+-k^+) \nonumber\\
	&\textrm{Tr}\left [\Gamma_{\lambda,\lambda'}\gamma^+ S_f(k_\eta)\Gamma^{\chi}(k;P)\{\cdot \epsilon_\Lambda(P)\}S_g(k_{\bar{\eta}})  \right ].
\end{align}
It works for both pseudoscalar and vector mesons. The $\Gamma^{\chi}(k;P)$ is the meson's BS amplitude, and the $S(k)$ is the dressed quark propagator. The $k_\eta=k+\eta P$ and $k_{\bar{\eta}}=k-(1-\eta)P$ are the momentum partitions, with $\eta$ set to $\frac{1}{2}$ in this work. The $\Gamma_{\pm,\mp}=I\pm \gamma_5$ and $\Gamma_{\pm,\pm}=\mp(\gamma^1\mp i\gamma^2)$ project out corresponding quark-antiquark helicity configurations \footnote{$\Gamma_{\pm,\mp}=I\pm \gamma_5$ refers to $\Gamma_{+,-}=I+\gamma_5$ and $\Gamma_{-,+}=I-\gamma_5$, which is by taking the sign in the same row simultaneously. This notation applies throughout this paper.}. The trace is taken over Dirac, color and flavor spaces. In this work, we take the u-d quark isospin symmetry. So we only consider $\pi^+(u\bar{d})$ and $\rho^+(u\bar{d})$ mesons without loss of generality. An implicit color factor $\delta_{ij}$ is associated with $\Gamma_{\lambda,\lambda'}$, along with $\sigma^-=\frac{1}{2}(\sigma_1-i \sigma_2)$ in flavor space. 

The $\Phi^{\{\Lambda\}}_{\lambda,\lambda'}(x,\vect{k}_T)$'s are not scalar functions, but they can be expressed with scalar amplitudes $\psi(x,\vect{k}_T^2)$'s \cite{Carbonell:1998rj,Ji:2003fw}.  For pseudoscalar meson, it takes the form 
\begin{align}
	\hspace{00mm}\Phi_{\pm,\mp}&=\psi_{(1)},  \ \ \ \ \ 
	&\Phi_{\pm,\pm}&=\pm k_T^{(\mp)} \psi_{(2)}, \label{eq:phi1}
\end{align}
and for vector meson
\begin{align}
	\hspace{00mm}\Phi_{\pm,\mp}^{0}&=\psi^{0}_{(1)},  \ \ \ \ \ 
	&\Phi_{\pm,\pm}^{0}&=\pm k_T^{(\mp)} \psi^{0}_{(2)}, \label{eq:phi1}\\
	\Phi_{\pm,\pm}^{\pm 1}&=\psi^{1}_{(1)},
	&\Phi_{\pm,\mp}^{\pm 1}&=\pm  k_T^{(\pm)}\psi^{1}_{(2)}, \notag \\
	\Phi_{\mp,\pm}^{\pm 1}&=\pm k_T^{(\pm)}\psi^{1}_{(3)},
	&\Phi_{\mp,\mp}^{\pm 1}&=(k_T^{(\pm)})^2\psi^{1}_{(4)}. \label{eq:phi2}
\end{align}
with $k_T^{(\pm)}=k^x \pm i k^y$. Assuming the isospin symmetry in $\rho(u\bar{d})$, one further has \cite{Shi:2021taf}
\begin{align}\label{eq:psi2}
	\psi^{1}_{(2)}(x,\vect{k}_T^2)&=-\psi^{1}_{(3)}(1-x,\vect{k}_T^2).
\end{align}

Using these $q\bar{q}$-LFWFs, we can calculate their contribution to the unpolarized PDFs of the mesons through the overlap representation 
\begin{align}
f_{q\bar{q}}^{\{\Lambda\}}(x)&=\sum_{\lambda,\lambda'}\int \frac{d \vect{k}_T^2}{2(2 \pi)^3}  |\Phi^{\{\Lambda\}}_{\lambda,\lambda'}(x,\vect{k_T})|^2. \label{eq:N2}
\end{align}
Consider its zeroth moment $N^{\{\Lambda\}}_{q\bar{q}}=\int_0^1 dx f^{\{\Lambda\}}_{q\bar{q}}(x)$, we find the it ranges between 30\% to 50\% for pion and $\rho$ mesons \cite{Shi:2021taf,Shi:2021nvg} \footnote{DSE studies in Minkowski space \cite{dePaula:2020qna} yields qualitatively same result, i.e., $N_{q\bar{q}}<1$.}. We pointed out this is a strong indication of complex higher Fock-components in light hadrons. These $q\bar{q}$-PDFs will be compared with the complete PDFs effectively incorporating all Fock-components in Sec.~\ref{sec:results}.

\section{Full PDFs of pion and $\rho$ within RL-DSEs.}\label{sec:DSE}
Within the DSE formalism, it is possible to compute PDFs that effectively incorporate contributions from all Fock states, thereby circumventing the light-front Fock-state expansion. We refer to such PDFs as full PDFs to distinguish them from $q\bar{q}$-PDFs, and simply as PDFs when no confusion arises. 

The full PDFs are obtained by directly evaluating covariant Feynman diagrams based on the operator definition of PDFs, rather than through the calculation of LFWFs. This approach was pioneered in \cite{Bednar:2018mtf}, which presented a full rainbow--ladder DSE calculation of the pion's unpolarized PDF. The central idea was to resum the gluon-ladder dressing effects of the probing light-cone vertex $\gamma^+\delta(x P^+-k^+)$. Building on this idea, we lately developed a new computational technique in which the gluon ladders are resummed into the hadron's quark--quark correlation matrix, from which various PDFs can be extracted.

\subsection{DSE of pseudoscalar meson's quark--quark correlation matrix}
  \begin{figure}[htbp]
	\centering
	\includegraphics[width=3.3in]{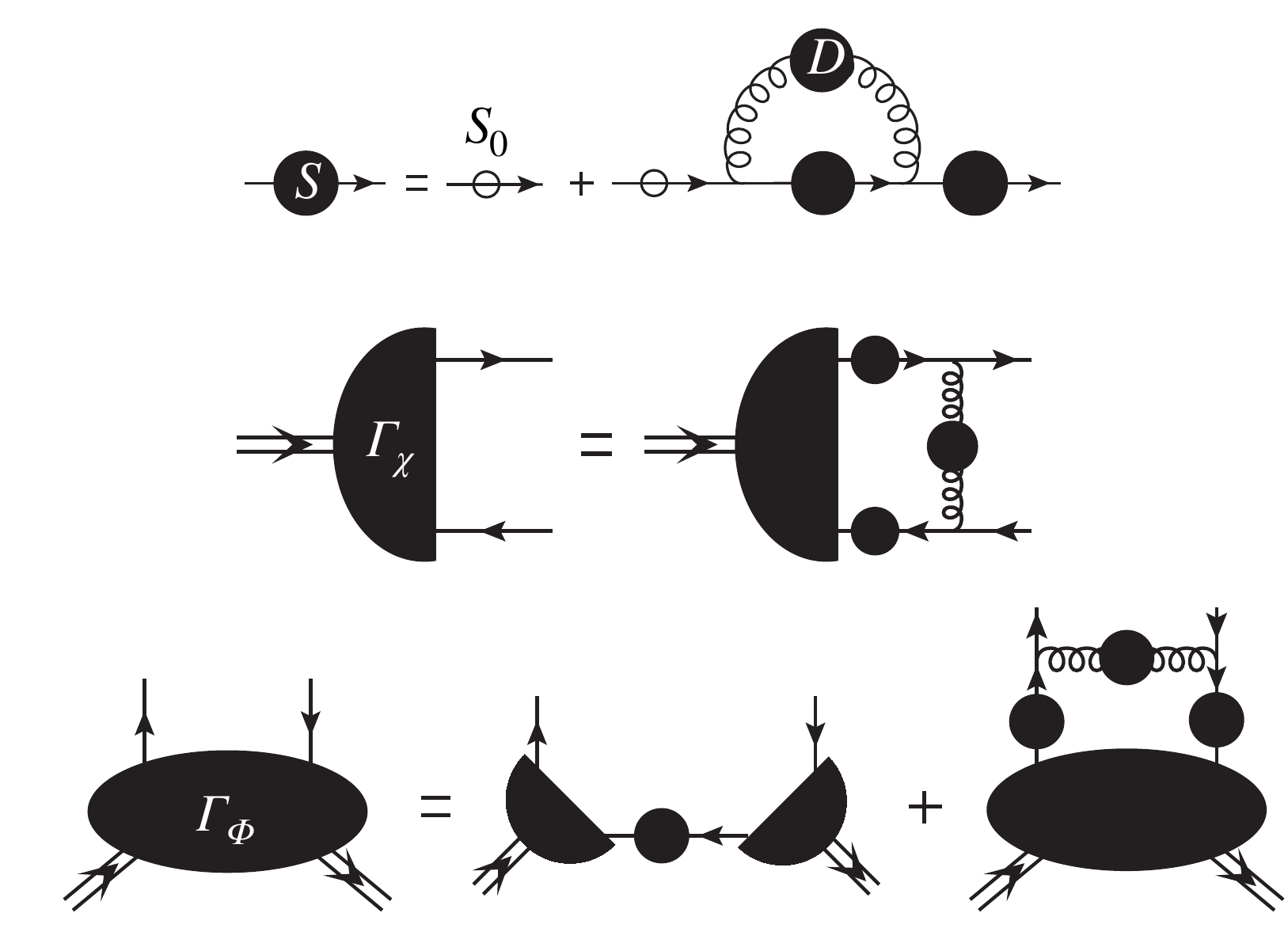}
	\caption{From top to bottom are the rainbow--ladder truncated DSEs for dressed quark propagator $S(k)$, the amputated Bethe--Salpeter bound state vertex $\Gamma_{\chi}(k;P)$, and the amputated quark--quark correlation matrix  $\Gamma_\Phi(k;P)$ respectively.}
	\label{fig:DSE}
\end{figure}
The RL-DSE of pseudoscalar meson's quark--quark correlation matrix in Euclidean space had been introduced in our study \cite{Shi:2026wca}, which will be recapitulated below. In Fig.~\ref{fig:DSE}, the DSEs under rainbow--ladder truncation for dressed quark propagator $S$, the amputated Bethe--Salpeter bound state vertex for the pion $\Gamma_\chi$, and the amputated pion quark--quark correlation  matrix  $\Gamma_\Phi$ are displayed from top to bottom respectively. In formulas, these RL-DSEs read 
\begin{align}
S(k)^{-1} &= Z_2 \,(i \kslash + Z_4 m_{\rm q}(\mu)) \nonumber \\ 
		  &\hspace{10mm}+ Z_2^2\!\! \int^{\Lambda_q}_q\!\! g^2 D_{\mu\nu}(q)
\frac{\lambda^a}{2}\gamma_\mu S(k-q) \frac{\lambda^a}{2}\gamma_\nu~, \label{eq:quarkDSE}\\
\Gamma^{\chi}(k;P) &=\!-Z_2^2 \!\!\int_q^{\Lambda_q}\!\!
\!\!g^2 D_{\mu\nu}(k\!-\!q) \,\gamma^{a}_{\mu} \,S(q_{\eta})\Gamma^{\chi}(q;P) S(q_{\bar{\eta}})\, \gamma^{a}_{\nu}~, \label{eq:mesonBSE}\\
\Gamma^\Phi(k;P) &=\Gamma^\chi (k;P)\, S(k_{\bar{\eta}})\,\bar{\Gamma}^\chi(k;-P) \nonumber\\
&-Z_2^2\!\! \int_q^{\Lambda_q}\!\!
g^2 D_{\mu\nu}(k\!-\!q) \gamma^{a}_{\mu}  S(q_\eta)\Gamma^\Phi(q;P) S(q_\eta) \gamma^{a}_{\nu} ~,\label{eq:phiDSE}
\end{align}
Here $\int^{\Lambda_q}_q$ represents 
\mbox{$\int d^4 q/(2 \pi)^4$} with  a smooth Poincar\'e invariant ultraviolet regularization at  mass-scale $\Lambda$.  The $m_{\rm q}(\mu)$ is the current quark mass renormalized at scale $\mu$. The $Z_{2}$ and $Z_4$ are the quark field and mass renormalization constants respectively. The factor $Z_2^2$ preserves multiplicative renormalizability in solutions of the DSE and BSE~\cite{Bloch:2002eq}.  The $D_{\mu\nu}$ is in  Landau gauge and it models a combination of quark-gluon vertex and gluon propagator~\cite{Maris:1997tm,Maris:2003vk}. Here we take the infrared part of Qin-Chang model \cite{Qin:2011dd}
\begin{equation}
\label{eq:QC}
D_{\mu\nu}(q) = \left(\delta_{\mu\nu}-\frac{q_\mu q_\nu}{q^2}\right)\frac{8 \pi^2}{\omega^4} D  \, {\rm e}^{-q^2/\omega^2}.
\end{equation} 
With this super-renormalizable model, the $\Lambda_q$ can be safely taken to infinity and renormalization constants $Z_2=Z_4=1$. Model parameters involved are current quark mass $m_q=5.5$ MeV, $\omega=0.5$ GeV and $D=1.1$ GeV$^{-2}$ \cite{Shi:2021taf,Shi:2021nvg}.

The $S$, $\Gamma^\chi$, $\Gamma^\Phi$ are essentially matrices with two Dirac indices. Their general Dirac structure decomposition are as follows: 
\begin{align}
S(k)^{-1} &=i\sh{k}\, A(k^2) + B(k^2) \label{eq:SDe}\\
\Gamma^\chi(k;P)&=\gamma^5 \left(\sum_{i=1}^4 T^{\chi}_i \mathcal{F}_{\chi,i}\left(k^2,k\cdot P,P^2\right)\right)~,\label{eq:GammaChiDe}\\
\Gamma^\Phi(k;P)&=\sum_{i=1}^4 T^{\Phi}_i \mathcal{F}_{\Phi,i}\left(k^2,k\cdot P,P^2\right)~.\label{eq:GammaPhiDe}
\end{align}
with Dirac matrix covariants  $T^\chi=T^\Phi=\{\textrm{I}_4, \sh{P}, \sh{k}, [\sh{k},\sh{P}]\}$ consistent with parity, time reversal and charge conjugation~\cite{Maris:1997tm,Tangerman:1994eh}.  The $A, B$ functions as well as the four  amplitudes ${\cal F}_{\chi/\Phi}=\{E_{\chi/\Phi}, F_{\chi/\Phi}, G_{\chi/\Phi}, H_{\chi/\Phi}\}$ are Lorentz scalars. Note that the BS equation (\ref{eq:mesonBSE}) is homogeneous, so the obtained BS amplitude has an ambiguous overall factor. In theory, this can be unambiguously determined by imposing physical requirement that the four point Green function has a pole with residue one around this bound-state, i.e.,
\begin{align}
G_4\propto \frac{\chi \bar{\chi}}{P^2-m_\pi^2}+{\rm regular}.	
\end{align}
This is also equivalent to the physical demand that the pion's valence quark number sums up to unity, or its electromagnetc form factor satisfies $F_\pi(Q^2=0)=1$. Under rainbow--ladder truncation, this leads to
\begin{align}
2 P_\mu &=\frac{\partial}{\partial P_{\mu}}\int_k^{\Lambda_k} {\rm Tr}_{\rm D, c, f}\left[\bar{\Gamma}^{\chi}(k;-K) S(k_\eta)\right. \nonumber \\
&\hspace{20mm}\times\left. \Gamma^{\chi}(k;K)S(k_{\bar{\eta}})\right]|_{P^2=K^2=-m^2} ,\label{eq:GammaChiNorm}
\end{align}
with $\bar{\Gamma}^{\chi}(k,-P)^T=C^{-1}\Gamma^{\chi}(-k,-P)C$ and $C=\gamma_2 \gamma_4$.

To numerically solve these equations, we proceed sequentially: first Eqs. (\ref{eq:quarkDSE}, \ref{eq:QC}, \ref{eq:SDe}); then Eqs.~(\ref{eq:mesonBSE}, \ref{eq:QC}, \ref{eq:GammaChiDe}); followed by Eq. (\ref{eq:GammaChiNorm}); and finally Eqs. (\ref{eq:QC}, \ref{eq:phiDSE}, \ref{eq:GammaPhiDe}). Finally we obtain the unamputated quark--quark correlation matrix $\Phi$ by attaching the dressed quark legs
\begin{align}
\Phi(k;P)=S(k_\eta)\Gamma^\Phi(k;P) S(k_\eta).	
\end{align}
 Note the $\Phi$ has same Dirac structure decomposition as Eq.~(\ref{eq:GammaPhiDe}).

\subsection{DSE of vector meson's quark--quark correlation matrix} 
In this section, we present the RL-DSE for the vector meson's quark--quark correlation matrix, which constitutes the central novelty of this work. Consider the unamputated and connectted quark--quark correlation matrix for the  $\rho$, which is defined as~
\begin{align} \label{eq:phidef}
\epsilon_{\mu,(\Lambda)}(P)\Phi^{\mu\nu}_{ij}(k,P)\epsilon^*_{\nu,(\Lambda)}(P)&=\nonumber \\
&\hspace{-20mm}\int \textrm{d}^4 \xi \ e^{ik_\eta \cdot \xi}\ {}_{\vect{n}}\langle P,\Lambda|\bar{\psi}_{j}(0)\psi_{i}(\xi)|P,\Lambda\rangle_{\vect{n}},	
\end{align}
The $\vect{n}$ is spin quantization axis, and $\epsilon_{\mu,(\Lambda)}(P)$ is the polarization 4-vector of vector meson for the state $|P,\Lambda\rangle_{\vect{n}}$. For our purpose of studying the unpolarized quark PDFs, we assign $\vect{n}$ to be aligned with $z-$direction.

Analogous to the pion case, the BS equations of vector meson's BS amplitude and $\Phi_{\mu\nu}$ read
\begin{align}
\Gamma^{\chi}_{\mu}(k;P) &=\!-Z_2^2 \!\!\int_q^\Lambda\!\!
\!\!g^2 D_{\rho\sigma}(k\!-\!q) \,\gamma^{a}_{\rho} \,S(q_{\eta})\Gamma^{\chi}_{\mu}(q;P) S(q_{\bar{\eta}})\, \gamma^{a}_{\sigma}~ \!, \label{eq:VmesonBSE}\\
\Gamma^\Phi_{\mu\nu}(k;P) &=\Gamma^\chi_\mu (k;P)\, S(k_{\bar{\eta}})\,\bar{\Gamma}^\chi_{\nu}(k;-P) \nonumber\\
&-Z_2^2\!\! \int_q^\Lambda\!\!
g^2 D_{\rho\sigma}(k\!-\!q) \gamma^{a}_{\rho}  S(q_\eta)\Gamma^\Phi_{\mu\nu}(q;P) S(q_\eta) \gamma^{a}_{\sigma} ~ \!,\label{eq:VphiDSE}
\end{align}
and the BS amplitude normalized by \cite{Maris:1999nt}
\begin{align}
2P_\mu&=\frac{\partial}{\partial P_\mu}\frac{1}{3} \int_q^{\Lambda_q} \textrm{Tr}_{\rm D,c,f}\left[\bar{\Gamma}^{\chi}_{\nu}(k;-K)S(k_\eta)\right. \nonumber\\
&\hspace{10mm}\left. \times\Gamma_\nu^{\chi}(k;K)S(k_{\bar{\eta}})\right]|_{P^2=K^2=-m^2}\label{eq:VGammaChiNorm}	
\end{align}

The meson BS amplitude $\Gamma^\chi_\mu$  has the most general Dirac structure decomposition
\begin{align}  \label{eq:gammachidecomp} 
\Gamma^\chi_\mu(k;P)=\sum_{i=1}^8 T^{\chi}_{i,\mu} \mathcal{F}^{\chi}_i\left(k^2,k\cdot P,P^2\right),
\end{align}
where the $T^{\chi}_{i,\mu}$ are matrix product of two group of basis 
\begin{align}
T_{i,\mu}^{\chi}\in\{\textrm{I}_4, \sh{P}, \sh{k}, [\sh{k},\sh{P}]\} \otimes \{k^{(T)}_\mu, \gamma^{(T)}_\mu\}. \label{VGammaChiDe}
\end{align}
The $k^{(T)}_\mu=k_\mu-P_\mu\frac{k\cdot P}{P^2}$ and $\gamma^{(T)}_\mu=\gamma_\mu-P_\mu\frac{\slashed{P}}{P^2}$ are constructed to be orthogonal to $P_\mu$ to satisfy $P\cdot \Gamma^\chi=0$. 

The $\Gamma^\Phi_{\mu\nu}$ can be decomposed as
\begin{align}  \label{eq:gammaphidecomp}
\Gamma^\Phi_{\mu\nu}(k;P)=\sum_{i=1}^{20} T^{\Phi}_{i,\mu\nu} \mathcal{F}^\Phi_i\left(k^2,k\cdot P,P^2\right)~,
\end{align}
where the most complete basis are 
\begin{align}\label{eq:VGammaPhiDe}
	T_{i,\mu\nu}^{\Phi}&\in\{\textrm{I}_4, \sh{P}, \sh{k}, [\sh{k},\sh{P}]\} \nonumber \\ 
	&\otimes \{ \gamma^{(T)}_\mu \gamma^{(T)}_\nu, \gamma^{(T)}_\nu \gamma^{(T)}_\mu, \gamma^{(T)}_\mu k^{(T)}_\nu, k^{(T)}_\mu \gamma_\nu^{(T)}, k^{(T)}_\mu k^{(T)}_\nu \}.
\end{align}
Note another possible tensor basis $\delta_{\mu\nu}-\frac{P_\mu P_\nu}{P^2}$ is not independent from those in the second bracket of Eq.~(\ref{eq:VGammaPhiDe}). Numerically solving these equations follow analogous sequences as for pion, only at a larger computing cost. Finally, the vector meson's unamputated quark--quark correlation matrix can be obtained by 
\begin{align}
\Phi_{\mu\nu}(k;P)&=S(k_\eta)\Gamma^{\Phi}_{\mu\nu}(k;P)S(k_\eta)
\end{align}

\subsection{From quark--quark correlation matrix to unpolarized PDFs}
The unpolarized PDFs can be obtained from the quark--quark correlation matrix \cite{Jaffe:1996zw,Barone:2001sp}. For pion, it is
\begin{align}\label{eq:fdef}
f_\pi(x)&=\frac{1}{2}\int\frac{\textrm{d}^4 k}{(2\pi)^4}\delta(k_\eta\CDOT n - xP\CDOT n)\textrm{Tr}\left[\Phi(k,P)\slashed{n} \right],
\end{align}
and for vector meson of helicity $\Lambda$ it is
\begin{align}\label{eq:fdefV}
f^{\Lambda}_\rho(x)&=\frac{1}{2}\int\frac{\textrm{d}^4 k}{(2\pi)^4}\delta(k_\eta\CDOT n - xP\CDOT n) \nonumber\\
&\hspace{20mm}\times \textrm{Tr}\left[\Phi_{\mu\nu}(k,P)\slashed{n} \right]\epsilon_{\mu,(\Lambda)}(P)\epsilon^*_{\nu,(\Lambda)}(P),
\end{align}
We then numerically compute the Mellin moments and reconstruct the concerned PDFs. In practice, we compute the $2x-1$-moments 
\begin{align}\label{eq:mom}
\langle (2x-1)^m \rangle_f &=\int_{-1}^1 dx (2x-1)^m f(x) 
\end{align}

The results are presented in Table~\ref{tab:moms}. Based on these moments, the PDFs can be reconstructed. We parameterize all unpolarized PDFs using a unified functional form~\cite{Shi:2024laj}
\begin{align}
\label{eq:fitf}
f(x)= x^\alpha (1-x)^\beta (c_0+c_1 x)+d_0+d_1 x,
\end{align}
where $\alpha$, $\beta$, and $c_i$, $d_i$ are fitting parameters. Although this parametrization is not a rigorous representation of the original distributions, it is sufficiently flexible to accommodate a wide range of possible shapes. In particular, we deliberately introduce a $d_0+d_1 x$ term to allow for the possibility that the PDFs remain finite at $x=1$\footnote{This does not conflict with the requirement that $f(x)=0$ for $x>1$, since model studies such as the NJL model yield analytic results exhibiting a possible discontinuous drop at $x=1$.}. With Eq.~(\ref{eq:fitf}), the Mellin moments are reproduced with a mean squared deviation below 0.5\%. The fitted parameters are listed in Table~\ref{tab:moms}.

%=====================================================================================================================
\begin{table*}[htbp]
\begin{center}
\begin{tabular}{c@{\hspace{0.8cm}}c@{\hspace{0.8cm}}c@{\hspace{0.8cm}}c@{\hspace{0.8cm}}c@{\hspace{0.8cm}}c@{\hspace{0.8cm}}c@{\hspace{0.8cm}}c@{\hspace{0.8cm}}c@{\hspace{0.8cm}}c@{\hspace{0.8cm}}}
\hline
 $m$  & $0$ & $1$ & $2$ & $3$ & $4$ & $5$ & $6$ & $7$ & $8$ \\\hline
$\langle (2x-1)^m\rangle_\pi$ & 1.00  & -0.257 & 0.321 & -0.159  & 0.189 & -0.114 & 0.134 & -0.0899 & - \\
$\langle (2x-1)^m\rangle_{\rho^{|\Lambda|=1}}$ & 1.00 & -0.241 & 0.294 & -0.142 &  0.167 & -0.0995 & 0.115 & -0.0760 & 0.0868 \\
$\langle (2x-1)^m\rangle_{\rho^{\Lambda=0}}$ &  1.00 & -0.105 & 0.353 & -0.0893  & 0.221 & -0.0770 & 0.162 & -0.0674 & 0.129 \\\hline
\end{tabular}
\end{center}
\vspace*{-4ex}
\caption{Computed Mellin moments of unpolarized PDFs of pion and $\rho$ mesons with $|\Lambda|=0,1$.
\label{tab:moms}
}
\end{table*}

\begin{table}[h!]

\begin{center}
\begin{tabular*}%{llcccccccc}
{\hsize}
{
c@{\extracolsep{0ptplus1fil}}|
c@{\extracolsep{0ptplus1fil}}
c@{\extracolsep{0ptplus1fil}}
c@{\extracolsep{0ptplus1fil}}
c@{\extracolsep{0ptplus1fil}}
c@{\extracolsep{0ptplus1fil}}
c@{\extracolsep{0ptplus1fil}}
c@{\extracolsep{0ptplus1fil}}}\hline
 $\pi$  & $\alpha$ & $\beta$ & $c_0$ &$c_1$ & $d_0$ &$d_1$  \\\hline
$f^{\pi}(x)$ & 0 & 0.805 & 5.747  & -0.871  & -3.997 & 3.975  \\
$f^{\rho^{|\Lambda|=1}}(x)$ & 0 & 2.725 & -1.168  & 0.548  & 2.521 & -2.476  \\
$f^{\rho^{\Lambda=0}}(x)$ & 0.131 & 0.241 & -5.505  & 6.808  & 5.231 & -5.300 \\\hline
\end{tabular*}
\end{center}
\vspace*{-4ex}
\caption{Fitting parameters for the pion and $\rho$ PDFs in Eq.~(\ref{eq:fitf})  
\label{tab:para}
}
\end{table}

\section{Results}\label{sec:results}

\begin{figure*}[htbp]
  \centering
  \includegraphics[width=\textwidth]{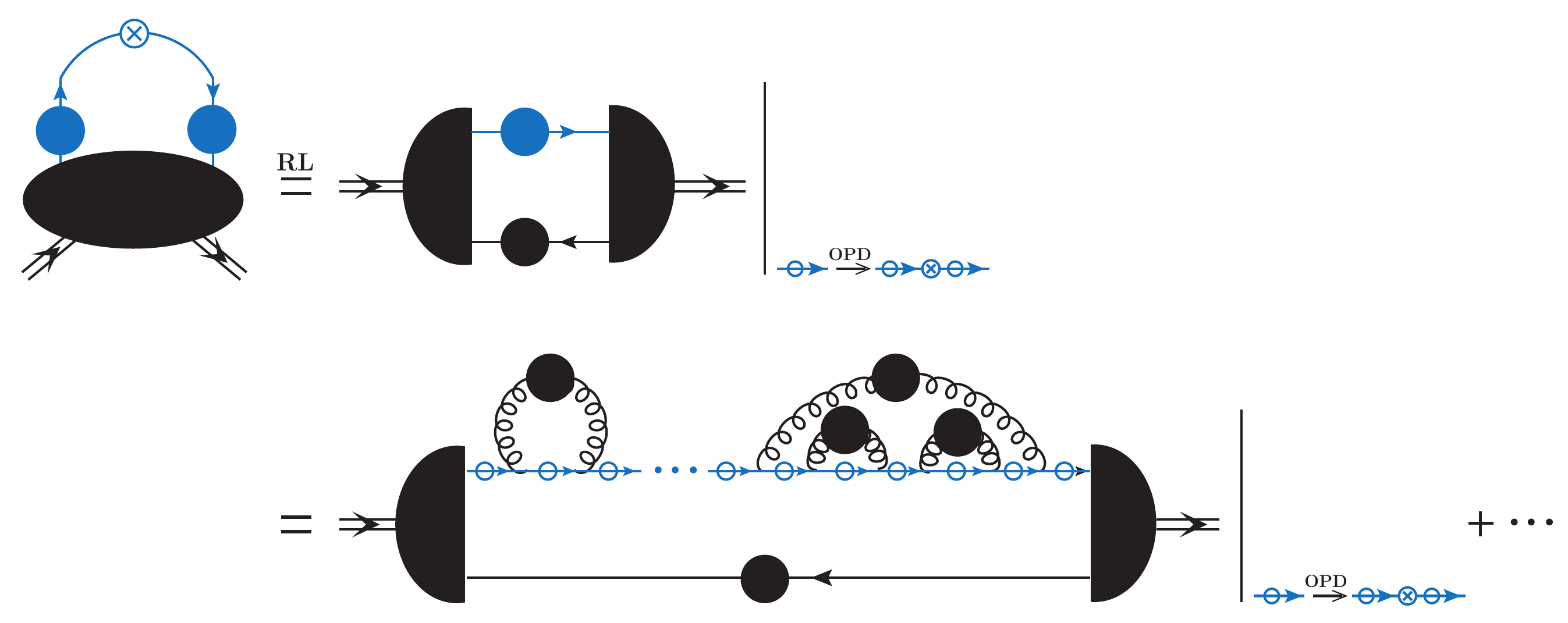}
  \caption{An intuitive picture of diagrams considered in RL-DSEs calculation of parton distribution function.}
  \label{fig:OPD}
\end{figure*}

Before showing the results of PDFs, we offer a viewpoint that may be helpful for understanding the diagrams included in our RL-DSE calculation of PDFs. Although DSEs are exact and nonperturbative, they can be motivated diagrammatically from a resummation of perturbative expansions. We adopt this diagrammatic viewpoint for its intuitive appeal. In this sense, the diagrams we consider in the calculation of Eq.~(\ref{eq:fdef}) or Eq.~(\ref{eq:fdefV}), are equivalent to a summation of infinitely many contributions, obtained by replacing one bare quark propagator $S_0(k_\eta)$ with
\[
S_0(k_\eta)\,\slashed{n}\,S_0(k_\eta)\,\delta(k_\eta \cdot n - x P \cdot n)
\]
within the struck dressed-quark propagator of the hadron-hadron transition amplitude, once per diagram (OPD); see the first line of Fig.~\ref{fig:OPD}. An example is shown in the second line of Fig.~\ref{fig:OPD}: after fully expanding the dressed propagator of the probed dressed quark in terms of bare propagators, this replacement is carried out for each $S_0$ once per diagram, thereby generating a large class of diagrams that add coherently. The black ellipsis denotes the infinitely many additional diagrams produced in this manner. Ultimately, every bare quark propagator in the transition amplitude is probed by the operator
\[
\slashed{n}\,\delta(k_\eta \cdot n - x P \cdot n),
\]
ensuring that all partonic contributions within the transition amplitude are systematically accounted for without double counting.

Apparently, the considered diagrams in Fig.~\ref{fig:OPD} only constitute a subset of all possible QCD diagrams, e.g., quark–antiquark pair creation and annihilation from the vacuum are ignored. This associates a model scale $\mu_0$ to our approach, at which intrinsic sea quark PDFs are absent, e.g., $f_{\bar{u}}(x;\mu_0)=f_{d}(x;\mu_0)=0$ in $\pi^+(u\bar{d})$ and $\rho^+(u\bar{d})$. Given $f_q(-x)=-f_{\bar{q}}(x)$, the obtained valence PDFs $f_{u/\bar{d}}(x)$ are thus only nonzero in the domain $x\in [0,1]$. Due to isospin-symmetry, only valence $u$ quark distribution in $\pi^+$ and $\rho^+$  will be considered in following, without loss of generality. We then evolve the PDFs using QCDNUM \cite{Botje:2010ay}. Next-to-leading order and variable flavor number schemes are taken, with  $\alpha_s(m_Z)=0.118$. The initial scale $\mu_0$ is determined to be $\mu_0=0.66$ GeV by matching to the JAM global fit result for pion $\langle x\rangle_{f_{\rm v,\rm u}}=0.268$ at 1.3 GeV \cite{Barry:2018ort}. 

  \begin{figure}[htbp]
	\centering
	\includegraphics[width=3.2in]{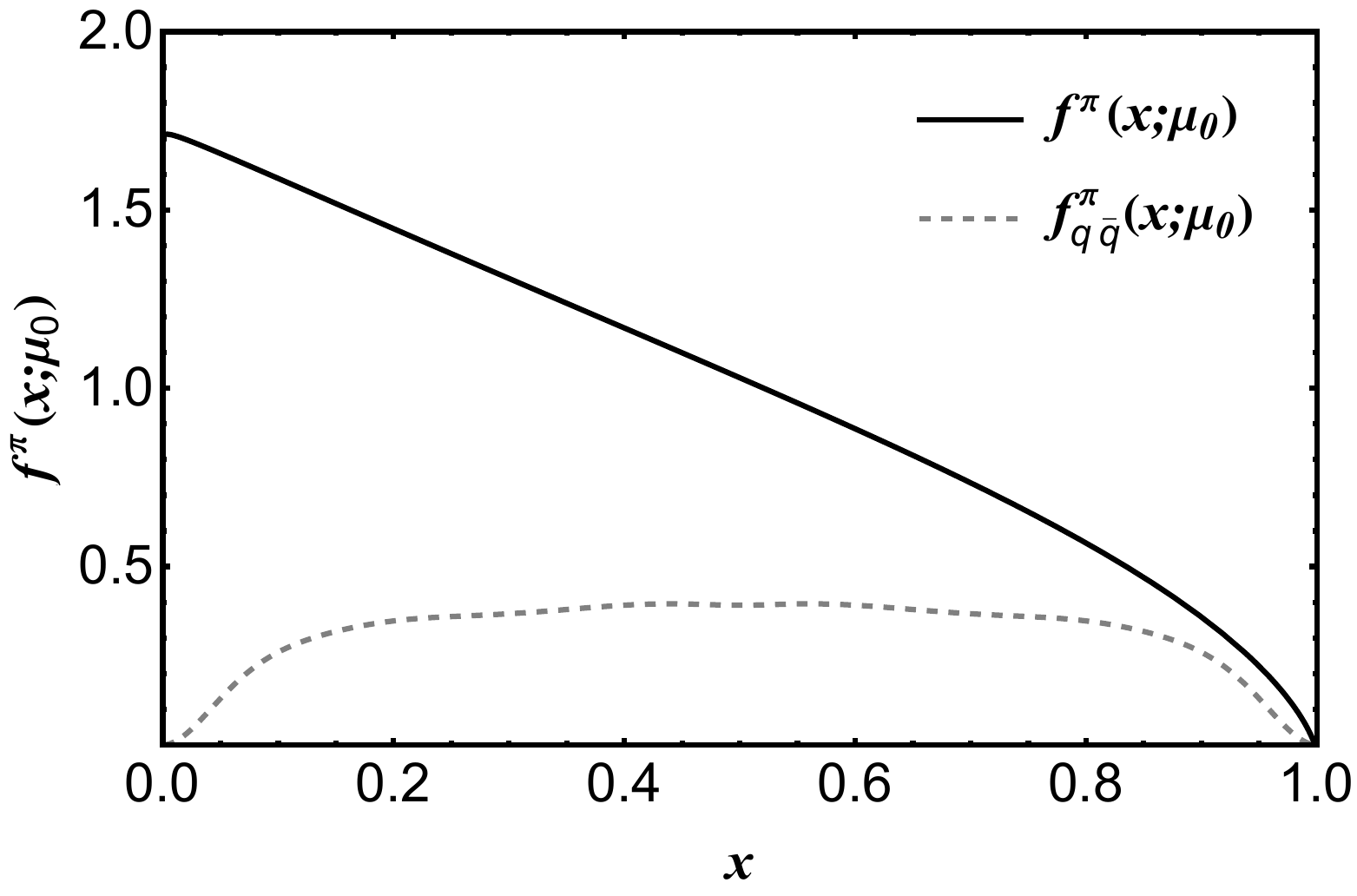}
		\includegraphics[width=3.2in]{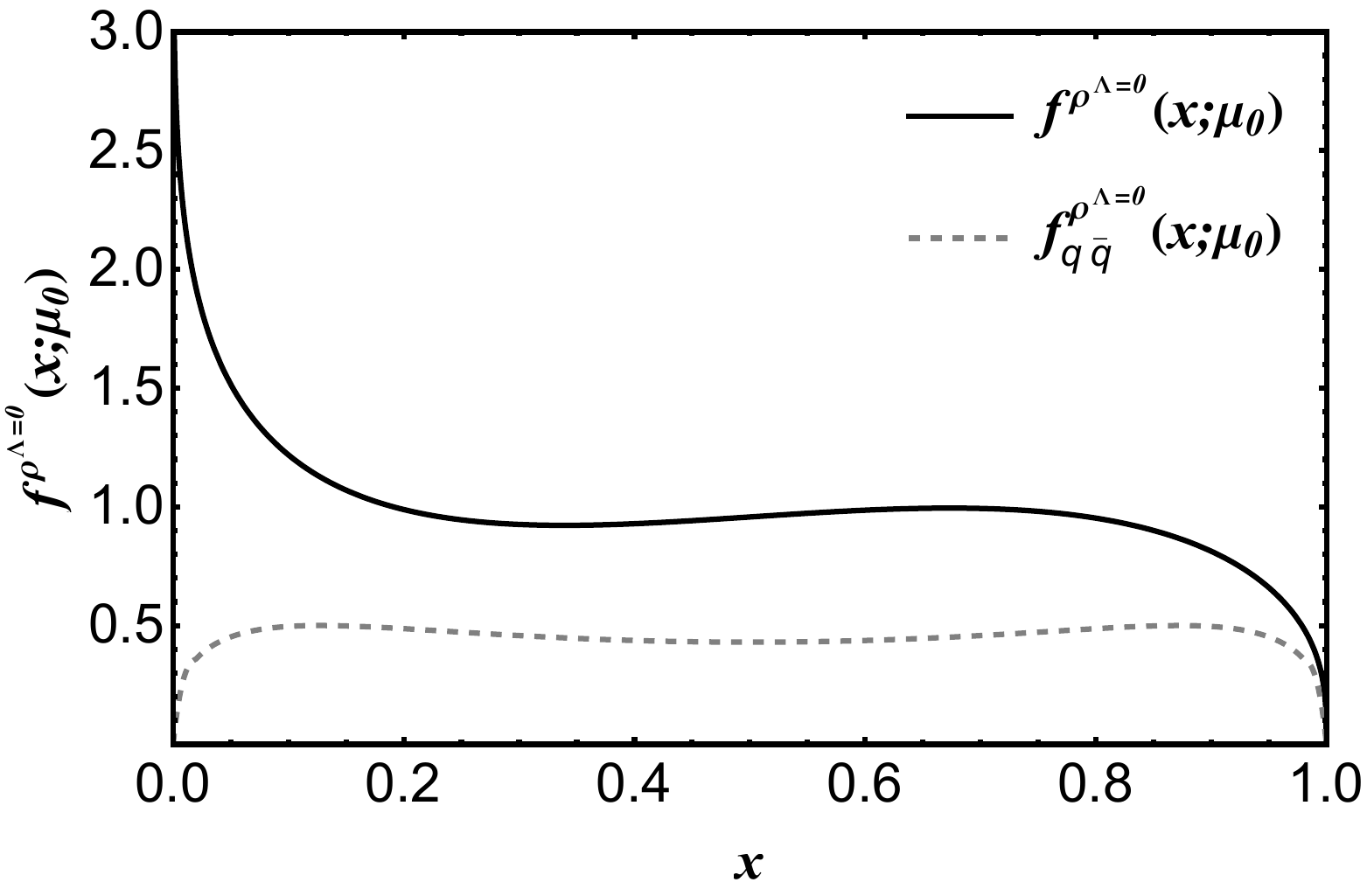}
			\includegraphics[width=3.2in]{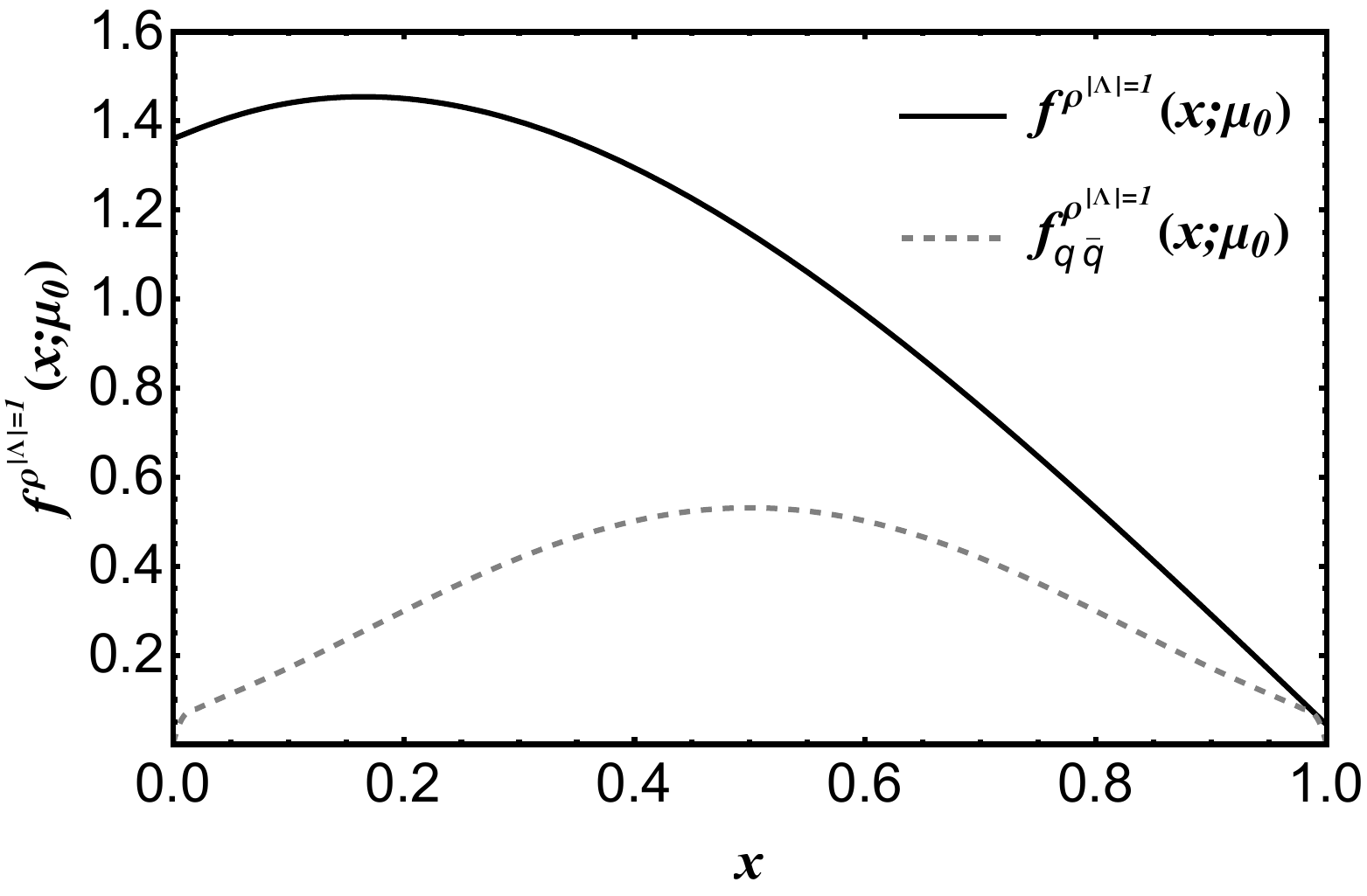}
	\caption{Pion and $\rho$'s unpolarized full PDFs and $q\bar{q}$-PDFs at hadronic scale.}
	\label{fig:PDF0}
\end{figure}

  \begin{figure}[htbp]
	\centering
	\includegraphics[width=3.5in]{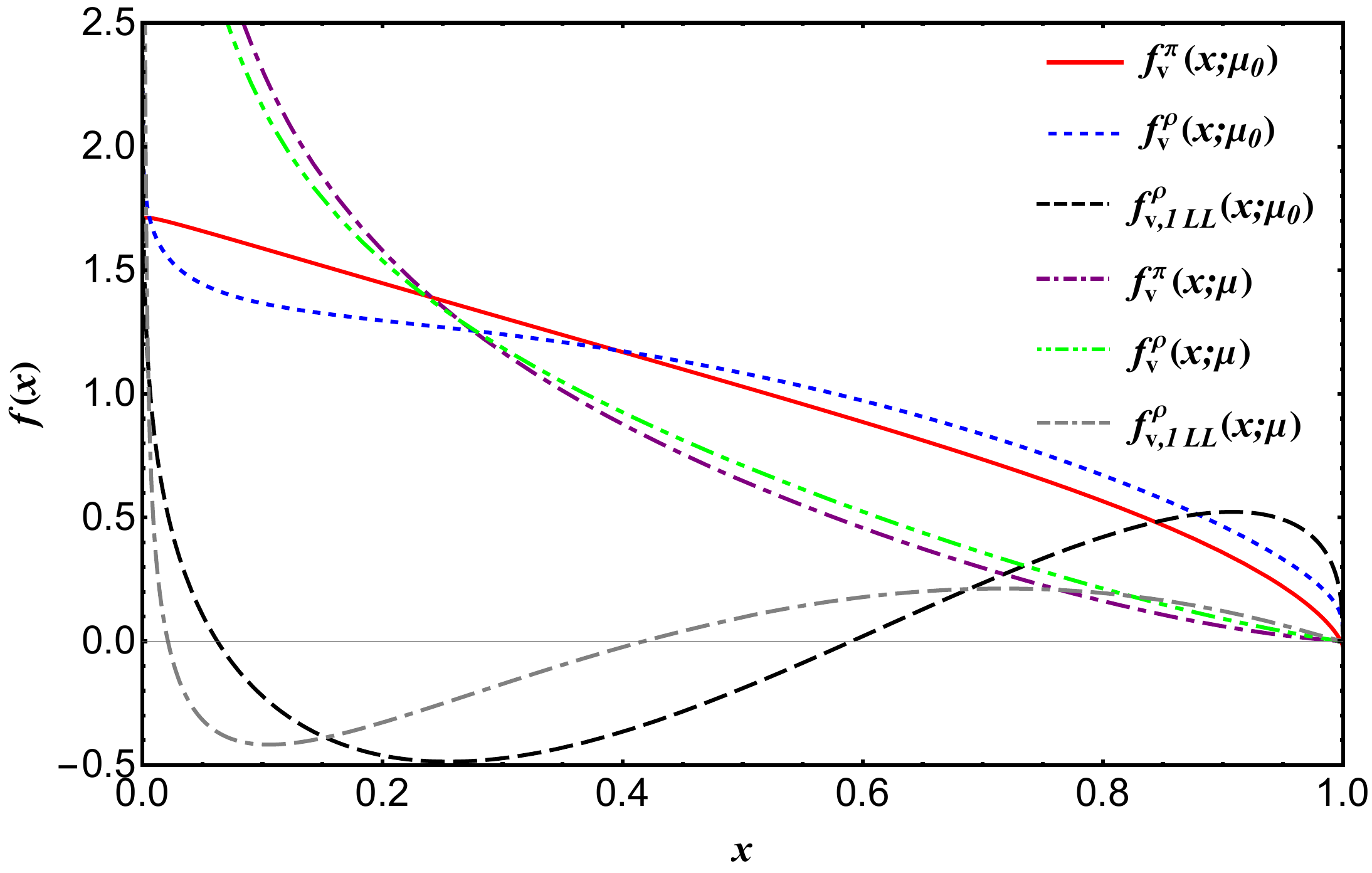}
	\caption{Pion and $\rho$'s valence PDFs at hadronic and experimental scales.}
	\label{fig:PDFH}
\end{figure}

The full unpolarized PDFs and $q\bar{q}$-PDFs of the pion and the $\rho$ at $\mu_0$ are displayed in Fig.~\ref{fig:PDF0}. Several common features can be observed across the three panels. First, all PDFs are skewed toward lower $x$, yielding $2\langle x \rangle_q=0.74, 0.76$, and $0.89$ for the $\pi$, $\rho^{\Lambda=\pm 1}$, and $\rho^{\Lambda=0}$, respectively. One may therefore infer that the remaining fraction of the meson's light-front momentum $P^+$ is carried by gluons, corresponding to $\langle x \rangle_g=0.26, 0.24$, and $0.11$, respectively. Second, the $q\bar{q}$-PDFs are significantly smaller than the corresponding full PDFs, and their profiles exhibit clear deviations, indicating that higher Fock states, such as $|u\bar{d}g\rangle$, which incorporate explicit gluon degrees of freedom, substantially reshape the PDFs. Notably, if one adopts the $q\bar{q}$ approximation, i.e., enforces $\langle q\bar{q} | q\bar{q} \rangle = 1$ by rescaling the $q\bar{q}$ LFWFs—substantial deviations from the full PDFs can arise. Third, the $q\bar{q}$-PDFs never exceed the full PDFs over the entire $x$ range, as required since contributions from higher Fock states to the PDFs are positive definite; this property is consistently preserved in our calculation. We emphasize that the $q\bar{q}$-PDFs were determined four years ago in our previous studies \cite{Shi:2021nvg,Shi:2021taf}, whereas the full PDFs presented here are obtained using exactly the same interaction model and parameters, without introducing or retuning any additional ones. Fourth, the $q\bar{q}$-PDFs predominantly contribute in the large-$x$ region; as $x$ decreases, higher Fock states are expected to play an increasingly important role.

On the other hand, several notable differences can be observed among these PDFs. The full PDFs of the $\rho$ are generally more skewed toward larger $x$, as is also reflected in their larger $\langle x \rangle$ values. This indicates that, at the hadronic scale, the pion contains a larger gluon momentum fraction than the $\rho$ mesons. Furthermore, the PDFs of $\rho^{\Lambda=0}$ and $\rho^{|\Lambda|=1}$ exhibit distinctly different profiles. This suggests that their difference, i.e., the tensor-polarized PDF (with unpolarized quarks) $f_{1LL}(x)$ of the $\rho$ meson, 
\begin{align}
f^\rho_{1LL}(x)\equiv\frac{1}{2}\left(2f^{\rho^{\Lambda=0}}(x)-\left(f^{\rho^{\Lambda=-1}}(x)+f^{\rho^{\Lambda=1}}(x)\right)\right)
\end{align}
 can be sizable, a point that will be examined below. In particular, the PDF of $\rho^{\Lambda=0}$ displays a novel structure: it develops a plateau in the valence region and shows a pronounced rise (while remaining finite at $x=0$) for $x \lesssim 0.1$.

We then compare the unpolarized PDFs of the pion, the $\rho$ meson
\begin{align}
f^\rho(x)\equiv\frac{1}{3}\left(f^{\rho^{\Lambda=-1}}(x)+f^{\rho^{\Lambda=0}}(x)+f^{\rho^{\Lambda=1}}(x)\right),
\end{align}
and the $f^{\rho}_{1LL}(x)$ at scales of $\mu_0$ and $\mu=2.4$ GeV. We consider the valence PDFs that are flavor non-singlet, e.g., $f_{\rm v,q}(x)\equiv f_{q}(x)-f_{\bar{q}}(x)$, and show them in  Fig.~\ref{fig:PDFH}. The unpolarized PDF at $\mu_0$ indicates that the valence quarks in the $\rho$ carry a larger fraction of light-front momentum than those in the pion \footnote{Our pion's valence quark PDF compares well with JAM global extraction \cite{Barry:2018ort}, as shown in \cite{Shi:2026wca}.}, although QCD evolution to higher scales gradually reduces this difference and brings the two distributions closer together. For the tensor-polarized PDF, given the sum rule $\langle x\rangle_{f_{1LL}}=0$ constraint and the strong enhancement of $f^{\rho^{\Lambda=0}}(x)$ at small $x$, the $f_{1LL}(x)$ develops a two-node structure, being positive, then negative, and positive again as $x$ increases. QCD evolution shifts these nodes toward smaller $x$. Somehow such pattern is opposite from most model studies, e.g., the NJL model \cite{Ninomiya:2017ggn,Zhang:2024nxl,Zhang:2024plq}, light front holographic model \cite{Kaur:2020emh}, BLFQ approach \cite{Kaur:2024iwn} and instanton liquid model \cite{Liu:2025fuf}. The studies that obtain results close to ours include a light-front quark model~\cite{Sun:2017gtz} and a simple wave-function model~\cite{Mankiewicz:1988dk}. Lattice QCD calculations have provided one or two moments of the $\rho$-meson PDFs~\cite{Best:1997qp,Loffler:2021afv}, but pointwise predictions for their $x$-dependence are not yet available. Interestingly, we note that the HERMES extraction of the deuteron $f_{1LL}(x)$ exhibits a positive signal at small $x$ that turns negative in the moderate-$x$ region~\cite{HERMES:2005pon}, qualitatively resembling the pattern we predict for the $\rho$ meson, albeit with a much smaller magnitude than our $\rho$ result \cite{Miller:2013hla,Kumano:2016ude,Cosyn:2017fbo}.

\section{Conclusion}\label{sec:con}
We present the first rainbow–-ladder Dyson–-Schwinger study of the vector meson’s quark–quark correlation matrix. Together with the pion case, we compare the resulting full unpolarized PDFs with the corresponding $q\bar{q}$-PDFs obtained from the $q\bar{q}$-LFWFs. A substantial gap is observed between the full PDFs and the $q\bar{q}$-PDFs, indicating the existence of significant intrinsic gluonic Fock components within our framework. We also find a pronounced difference between the PDFs of transversely ($|\Lambda|=1$) and longitudinally ($\Lambda=0$) polarized $\rho$ mesons. As a consequence, the tensor-polarized PDF of the $\rho$ is sizable in magnitude. It exhibits a novel two-node structure, being positive, then negative, and positive again as $x$ increases.

This work provides a further support for our argument that the RL-DSE incorporating gluons contains gluonic Fock-state component at parton level \cite{Shi:2021nvg,Shi:2021taf}, now substantiated through a detailed analysis of PDFs. It also demonstrates the strength of present approach in circumventing the Fock-state expansion and effectively resumming contributions from all Fock sectors in the investigation of parton distribution functions. Such a framework can also be extended to the spin-$1/2$ nucleon system in an analogous manner in the future.

\noindent\textbf{Acknowledgement:} 
We thank valuable suggestions from Qin-Tao Song.

\bibliography{pirhoPDF}

\end{document}